\documentclass[prl,twocolumn,preprintnumbers,amsmath,amssymb]{revtex4}

\usepackage{graphicx}
\usepackage{dcolumn}
\usepackage{bm}
\usepackage{color}
\usepackage{epstopdf}

\newcommand{\br}{ {\bm r}}

\def\bbeta{{\boldsymbol{\beta}}}
\def\bsigma{{\boldsymbol{\sigma}}}

\begin{document}



\title{Dramatic acceleration of wave condensation mediated by disorder in multimode fibers}

\author{Adrien Fusaro$^{1}$, Josselin Garnier$^{2}$, Katarzyna Krupa$^{3,1}$, Guy Millot$^{1}$, Antonio Picozzi$^{1}$}
\affiliation{$^{1}$ Laboratoire Interdisciplinaire Carnot de Bourgogne, CNRS, Universit\'e Bourgogne Franche-Comt\'e, Dijon, France}
\affiliation{$^{2}$ Centre de Mathematiques Appliqu\'ees, Ecole Polytechnique, 91128 Palaiseau Cedex, France}
\affiliation{$^{3}$ Dipartimento di Ingegneria dell'Informazione, Universit\`a di Brescia, via Branze 38, 25123, Brescia, Italy}


\begin{abstract}
Classical nonlinear waves exhibit a phenomenon of condensation that results from the natural irreversible process of thermalization,
in analogy with the quantum Bose-Einstein condensation.
Wave condensation originates in the divergence of the thermodynamic equilibrium Rayleigh-Jeans distribution, which is responsible for the macroscopic population of the fundamental mode of the system. 
However, achieving complete thermalization and condensation of incoherent waves through nonlinear optical propagation is known to require prohibitive large interaction lengths.
Here, we derive a discrete kinetic equation describing the nonequilibrium evolution 
of the random wave in the presence of a structural disorder of the medium.
Our theory reveals that a weak disorder accelerates the rate of thermalization and condensation by several order of magnitudes.
Such a counterintuitive dramatic acceleration of condensation provides a natural explanation for the recently discovered phenomenon of optical beam self-cleaning.
Our experiments in multimode optical fibers report the observation of the transition from an incoherent thermal distribution to wave condensation, with a condensate fraction of up to 60\% in the fundamental mode of the waveguide trapping potential. 
\end{abstract}


\maketitle

{\it Introduction.-}
The observation of the phenomenon of Bose-Einstein condensation has been reported in a variety of genuine quantum systems, such as ultracold atoms and molecules \cite{stringari}, exciton polaritons \cite{carusotto13} and photons \cite{weitz}. 
On the other hand, recent studies on wave turbulence revealed that a purely classical system of random waves can exhibit a process of condensation with thermodynamic properties analogous to those of Bose-Einstein condensation \cite{Newell01,nazarenko11,Newell_Rumpf,
PRL05,berloff07,PD09,Fleischer,suret,magnons15,nazarenko16,PRL18}.
Classical wave condensation finds its origin in the natural thermalization of the wave system toward the 
Rayleigh-Jeans equilibrium distribution, whose divergence is responsible for the macroscopic occupation of the fundamental mode of the system \cite{Newell01,nazarenko11,Newell_Rumpf,
PRL05,zakharov92,shrira_nazarenko13,chiocchetta16,PR14,laurie12}.
This self-organization process takes place in a formally reversible system:
The formation of the coherent structure (`condensate') remains
immersed in a sea of small-scale fluctuations (`uncondensed particles'), which store the information for time reversal.

There is a current surge of interest in studying quantum properties of fluids with light waves, such as superfluidity and the generation of Bogoliubov sound waves \cite{carusotto13,vocke16,michel18}.
Along this way, different forms of condensation processes have been reported in optical cavity systems, which are inherently nonequilibrium forced-dissipative 
systems \cite{PR14,conti08,fischer14,turitsyn13}.
On the other hand, the irreversible process of condensation is predicted for purely {\it conservative and formally reversible} (Hamiltonian) systems of random waves.
Unfortunately, however, the experimental study of condensation in a conservative (cavity-less) configuration constitutes a major challenge, because of the prohibitive large propagation lengths required to achieve thermalization \cite{chiocchetta16,PRL18}. 
In marked contrast with this commonly accepted opinion, an astonishing phenomenon of spatial beam self-organization, termed `beam self-cleaning', has been recently discovered in multimode optical fibers (MMFs) \cite{krupa16,wright16,krupa17}.
This phenomenon 
is due to a purely conservative Kerr nonlinearity \cite{krupa17} and, so far, its underlying mechanism remains unexplained.

As a matter of fact, light propagation in MMFs is known to be affected by a structural disorder of the material due to inherent imperfections and external perturbations \cite{kaminow13}, a feature of interest, e.g. in image formation \cite{psaltis16} or to study integrable nonlinear Manakov systems \cite{mecozzi12a,mecozzi12b,mumtaz13,xiao14}.
The remarkable result of our work is to show that a (`time'-dependent) structural disorder is responsible for a dramatic acceleration of the process of wave condensation.
On the basis of the wave turbulence theory \cite{Newell01,nazarenko11,zakharov92} and related recent developments on finite size effects \cite{Lvov10,Harris13,Harper13,Bustamante14}, we formulate a nonequilibrium kinetic description of the random waves that accounts for the impact of disorder.
The theory reveals that a {\it conservative} disorder introduces an effective dissipation in the system,
which is shown to deeply modify the regularization of resonant wave interactions.
We derive a {\it discrete} kinetic equation revealing that a weak disorder accelerates the rate of thermalization and condensation by several order of magnitudes.
Note that at variance with the notion of prethermalization to out of equilibrium states \cite{PRL18,Ruffo2009,Schmiedmayer16}, 
here the system achieves a fast relaxation to a fully thermalized equilibrium state.
The counterintuitive mechanism of condensation acceleration can provide a natural explanation for the effect of optical beam self-cleaning.
Our experiments realized in MMFs report the first observation of wave condensation featured by a macroscopic population of the fundamental mode of a (cavity-less) waveguide trapping potential.




The present work contributes to the challenging question of spontaneous organization of coherent states in nonlinear disordered (turbulent) systems
\cite{leuzzi15,delre17,churkin15,silberberg17,segev,cherroret15,schirmacher18,ermann15}.
Furthermore, MMFs are attracting 
for telecommunication applications 
\cite{kaminow13} and 
novel fiber laser sources \cite{wright17}.

\begin{figure}[]
\begin{center}
\includegraphics[width=0.85\columnwidth]{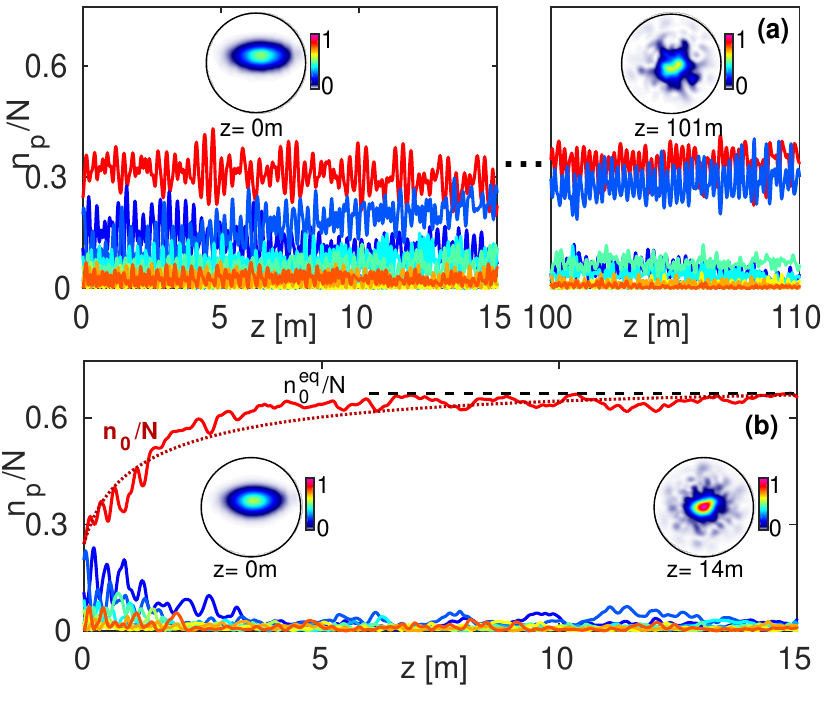}
\caption{\textbf{Disorder-induced condensation:} (a) Numerical simulations of the modal NLS Eq.(\ref{eq:mode_scalar}) in the absence of structural disorder (a), and in the presence of structural disorder (b) [NLS Eq.(\ref{eq:nls_Ap})], starting from the same initial coherent condition.
Evolutions of the modal populations $n_p/N$: fundamental mode $p=0$ (red), $p=1$ (dark blue), $p=2$ (blue), $p=3$ (light blue), $p=4$ (cyan), $p=5$ (light green), $p=6$ (green), $p=7$ (yellow), $p=8$ (orange).
A structural disorder breaks the coherent modal interaction (a) and induces condensation in the fundamental mode $n_0$ (b), which relaxes to the theoretical value $n_0^{eq}/N = 0.67$ at thermal equilibrium (dashed black).
The dotted red line in (b) reports $n_0(z)$ from the simulation of the kinetic Eq.(\ref{eq:kin_np}).
The insets show the corresponding intensity patterns $|\psi|^2(\br,z)$ ($N_*=120$ modes without including polarization degeneracy).
}
\label{fig:1}
\end{center}
\end{figure}

{\it Nonlinear Schr\"odinger model.-}
We consider the standard scalar (2+1)D nonlinear Schr\"odinger (NLS) equation,
which is known to describe the propagation (along $z$) of a polarized optical beam in a waveguide modelled by a confining potential $V(\bm r)$ (with $\bm r=(x,y)$) \cite{horak12}.
The potential $V(\bm r)$ is parabolic-shaped, which models graded-index MMFs  \cite{PR14,krupa16,wright16,krupa17,kaminow13}, or trapped Bose-Einstein condensates \cite{stringari}.
We expand the random wave into the basis of the linear eigenmodes $u_p(\br)$ (with eigenvalues $\beta_p$) of the $N_*$ modes of the fiber, 
${\psi}(\br,z)=\sum_{p} {A}_{p}(z) u_p(\br)$.
The modal components $A_p(z)$ of the NLS equation verify 
\begin{eqnarray}
i \partial_z {A}_p = \beta_p {A}_p - \gamma  F_p({A}),
\label{eq:mode_scalar}
\end{eqnarray}
where the modal expansion of the cubic Kerr nonlinearity reads $F_p({A})= \sum_{q,l,m} S_{pqlm} A_q A_l A_m^*$ and the tensor
$S_{pqlm}$ accounts for the spatial overlap among the eigenmodes \cite{horak12,PR14}.

We report in Fig.~\ref{fig:1}(a) a typical evolution of the modal components $A_p(z)$ and corresponding intensity pattern $|\psi|^2(\br,z)$, obtained by solving the NLS Eq.(\ref{eq:mode_scalar}) with  typical experimental parameters \cite{krupa17}, i.e., a MMF with $N_*=120$ modes (core radius $R=26\mu$m, $n_2=3.2\times 10^{-20}$m$^2$/W) and injected power $N=47.5$kW. 
At variance with usual simulations of wave turbulence \cite{nazarenko11,shrira_nazarenko13,PR14,nazarenko16}, we did not impose a random phase among the initial modes $A_p(z=0)$, which is consistent with the experimental conditions where a laser beam featured by a coherent transverse phase front is launched into the optical fiber.
The simulations of Eq.(\ref{eq:mode_scalar}) show that a strong phase-correlation among the modes is preserved during the propagation, thus leading to a phase-sensitive {\it coherent regime} of modal interaction.
The modes $A_p(z)$ then experience a quasi-reversible exchange of power with each other (Fig.~\ref{fig:1}(a)), which leads to an oscillatory dynamics of the intensity pattern $|\psi|^2(\br,z)$ (see the movie in \cite{SI}).
Such a multimode beam does not exhibit an enhanced brightness that characterizes a stable self-cleaning effect.
This coherent regime of mode interactions then freezes the thermalization process, a feature of growing interest that is analyzed in the framework of finite size effects in discrete or mesoscopic wave turbulence \cite{Lvov10,Harris13,Harper13,Bustamante14}.

\begin{figure}[]
\begin{center}
\includegraphics[width=.8\columnwidth]{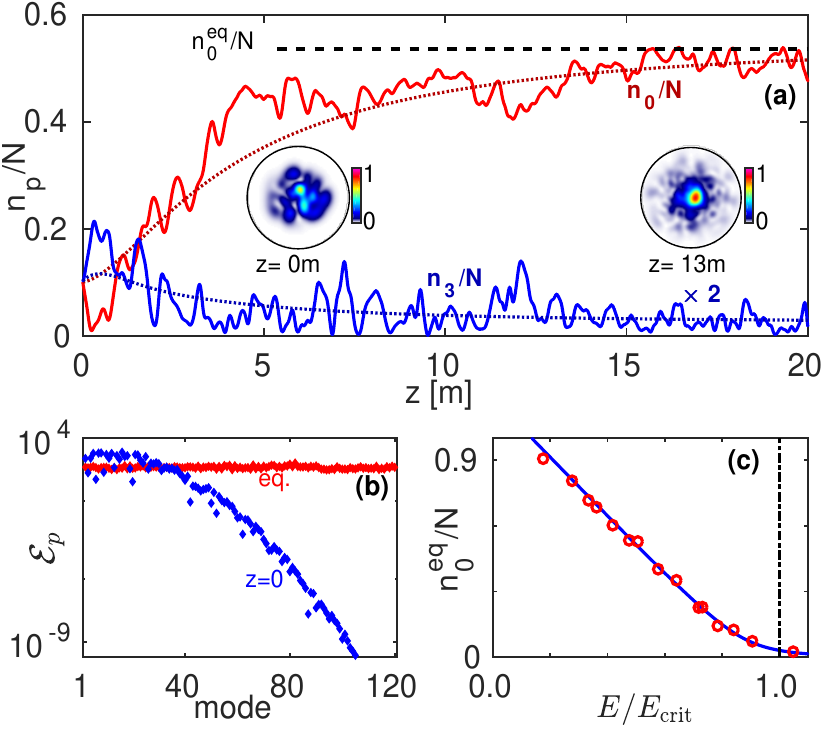}
\caption{\textbf{Condensation to equilibrium}.
(a) Simulation of the modal NLS Eq.(\ref{eq:nls_Ap}) (accounting for structural disorder) showing the evolution of $n_p(z)$ for the fundamental mode $p=0$ (red), and $p=3$ (blue), and corresponding simulation of the kinetic Eq.(\ref{eq:kin_np}) (dotted red and blue lines) starting from the same initial incoherent condition:  
$n_0(z)$ reaches the theoretical equilibrium value $n_0^{eq}/N \simeq 0.57$ (dashed black).
(b) Corresponding energy per mode ${\cal E}_p(z)=(\beta_p-\beta_0)n_p(z)$ at $z=0$ (blue dots), and at $z=20$m (red dots) showing that the beam reaches thermal equilibrium with {\it energy equipartition} ${\cal E}_p^{eq}=T$.
(c) Condensate fraction at equilibrium $n_0^{eq}/N$ vs energy $E$ (keeping constant the power $N$): For $E \le E_{\rm crit}$ the system undergoes a phase transition to condensation.
The NLS simulations of Eq.(\ref{eq:nls_Ap}) (red circles) are in quantitative agreement with the theory (blue line) without adjustable parameters.
}
\label{fig:2}
\end{center}
\end{figure}

{\it Kinetic equation with weak disorder.-}
Light propagation in MMFs is known to be affected by a structural disorder. We show below that disorder breaks the coherent modal regime discussed in Fig.~1(a) and leads to a turbulent incoherent regime with uncorrelated random phases fluctuations of the modes.
We stress that the acceleration of thermalization predicted by our theory is {\it not} simply due to a breaking of the coherent regime by disorder, but solely results from the interplay of disorder and the incoherent modal interaction.
Note that our theory goes beyond the mean-field approximation reported in \cite{krupa17}, which is a formally {\it reversible} theory inherently unable to explain the irreversible process of spatial beam self-cleaning.

Random mode coupling in MMFs has been widely studied in recent years \cite{kaminow13,mecozzi12a,mecozzi12b,mumtaz13,xiao14}.
Here, we consider the most general form of `space-time' disorder by introducing the random potential ${\hat {\bm W}}(\br,z)$ in the vector NLS equation 
$i \partial_z {\bm \psi} =-\alpha \nabla^2 {\bm \psi} + V(\bm r){\bm \psi}  -\gamma_0{\cal F}({\bm \psi}) + {\hat {\bm W}}(\br,z) {\bm \psi}$.
Its vector form accounts for the polarization degree of freedom of the field ${\bm \psi}({\bm r}, z)=(\psi_1, \psi_2)^T$ where ${\cal F}({\bm \psi})$ describes the corresponding cubic nonlinearity.
Because of the conservation of the power (`number of particles') $N=\int |{\bm \psi}|^2 d\br$, the random potential ${\hat {\bm W}}(\br,z)$ is Hermitian.
Therefore it can be expanded on the complete basis formed by the Pauli matrices $\hat{\bm W}(\br,z) =\sum_{j=0}^3 {\hat \nu}_j(\br,z) \bsigma_j$, where ${\hat \nu}_j(\br,z)$ are independent and identically distributed real-valued random processes, with variance $\sigma^2_\beta$ and correlation length $l_\beta$. 
The effective strength of disorder is controlled by the parameter $\Delta \beta = \sigma_\beta^2 l_\beta$.
We assume that disorder is a perturbation with respect to linear propagation $L_{lin}=\beta_0^{-1} \ll L_{disor}=1/\Delta \beta $ (with $\beta_0 l_\beta \gg 1$) and that it dominates nonlinear effects ($L_{disor} \ll L_{nl}$).
As for the modal NLS Eq.(\ref{eq:mode_scalar}), we expand the random wave into the eigenmodes of the linearized NLS equation without disorder, ${\bm \psi}(\br,z)=\sum_{p} {\bm A}_p(z) u_p(\br)$: 
\begin{eqnarray}
i \partial_z {\bm A}_p = \beta_p {\bm A}_p + {\bf W}_p(z) {\bm A}_p 
- \gamma  {\bm F}_p({\bm A}),
\label{eq:nls_Ap}
\end{eqnarray}
where the random matrices ${\bf W}_p(z)=\sum_{j=0}^3 \nu_{p,j}(z) \bsigma_j$  describe the dominant {\it weak disorder} contribution \cite{mumtaz13} and ${\bm F}_p({\bm A})$ the nonlinear coupling among the modes ($\gamma = \gamma_0 \int |u_0|^4(\br) d\br$) \cite{SI}. 
The modal NLS Eq.(\ref{eq:nls_Ap}) conserves the power $N=\sum_p |{\bm A}_p|^2(z)$ and the linear contribution to the energy $E=\sum_p \beta_p |{\bm A}_p|^2(z)$, which dominates the nonlinear contribution.
Indeed, in the experiments of beam self-cleaning the system evolves in the weakly nonlinear regime where linear dispersion (diffraction) effects dominate nonlinear effects $L_{lin}=\beta_0^{-1} \ll L_{nl}$.
Notice that since disorder is (`time') $z-$dependent, our system is of different nature than those studying the interplay of thermalization and Anderson localization \cite{cherroret15}.

In the limit of rapid disordered fluctuations, 
the generalized NLS equation has been reduced to the integrable Manakov equation \cite{mecozzi12a,mecozzi12b,mumtaz13,xiao14}, which, however, does not describe the process of beam self-cleaning.
Here, we go beyond the Manakov limit by deriving the following kinetic equation governing the evolution of the averaged modal components $n_p(z)=\left< |{\bm A}_p(z) |^2 \right>$ \cite{SI}:
\begin{eqnarray}
\nonumber
\partial_z n_p(z) &=&  \frac{ \gamma^2}{6\Delta \beta} \sum_{q,l,m}
\delta^K_{\beta_q+\beta_l- \beta_m - \beta_p}  |S_{pqlm}|^2 M_{pqlm}({\bm n}) \\
&&
+  \,  \frac{4\gamma^2}{9 \Delta \beta}  \sum_q  \delta^K_{\beta_q-\beta_p}
 |  s_{pq}({\bm n}) |^2 (n_q-n_p),
\label{eq:kin_np}
\end{eqnarray}
with $s_{pq}({\bm n})=\sum_{m'} S_{pqm'm'} n_{m'}$, and $M_{pqlm}({\bm n})=  n_q n_l n_p+n_q n_l n_m -  n_m n_p n_l -n_m n_p n_q$, where `$n_m$' stands for 
`$n_m(z)$', 
while the Kronecker symbol reflects energy conservation of wave resonances. 
The discrete nature of the kinetic equation originates in finite size effects due to the relative small number of modes of the trapping potential ($\sim$100 modes with $\beta_0 \gg 1/L_{nl}$) \cite{SI}.

{\it Acceleration of thermalization.-}
The originality of our theoretical approach with respect to the conventional wave turbulence approach \cite{Newell_Rumpf,nazarenko11,zakharov92} and the recent developments \cite{Lvov10,Harris13,Harper13,Bustamante14}, relies on the fact that it accounts for the presence of a structural disorder.
By using the Furutsu-Novikov theorem, our theory reveals that disorder deeply affects the evolution of the moments equations.
Specifically, the dynamics of a fourth-order moment of the random wave is governed by an effective forced-damped oscillator equation, in which the dissipation originates from the 
the conservative disorder.
It turns out that the singularity 
associated to a resonance is 
regularized by the dissipation due to disorder: {\it The lower the magnitude of disorder $\Delta \beta$, the stronger the efficiency of the wave resonance}.
Clearly, the amount of disorder $\Delta \beta$ cannot decrease arbitrarily since $\Delta \beta L_{nl} \gtrsim 1$ -- in the opposite regime $\Delta \beta \ll 1/L_{nl}$ the regularization due to disorder is negligible and the kinetic equation recovers the standard continuous form \cite{nazarenko11,PR14}.
The characteristic lengths (`times') of thermalization ($\zeta_{th}$) in the presence and the absence of disorder scale as 
\begin{eqnarray}
\zeta_{th}^{disor}/\zeta_{th}^{ord} \sim \Delta \beta / \beta_0.
\label{eq:ord_dis}
\end{eqnarray}
This shows that thermalization is significantly accelerated by the perturbative disorder $\Delta \beta/\beta_0 \ll 1$.
Considering typical experimental parameters \cite{krupa17} used in the simulations of Figs.~\ref{fig:1}-\ref{fig:2}, $\beta_0 \simeq 5 \times 10^{3}$m$^{-1}$, $L_{disor} \simeq 0.4$m (beat length $2\pi/\sigma_\beta \simeq 2.14$m, $l_\beta \simeq 30$cm), 
one obtains $\Delta \beta /\beta_0 \lesssim  5 \times 10^{-4}$:
The rate of thermalization is increased by several orders of magnitude by the disorder. 
Although the experimental parameters of disorder are not precisely known, the effect of disorder induces beam cleaning shown in Figs.~\ref{fig:1}-\ref{fig:2} is robust and has been observed over a {\it wide range of parameters} (e.g., $2\pi/\sigma_\beta$ and $l_\beta$ of several meters) \cite{SI}.

The kinetic Eq.(\ref{eq:kin_np}) 
describes a process of wave condensation \cite{PRL05,nazarenko11,PR14}.
It conserves the `number of particles' $N$, the energy $E$ and exhibits a $H-$theorem of entropy growth for the nonequilibrium entropy ${\cal S}(z)=\sum_p n_p(z) \log\big(n_p(z)\big)$, so that it describes an irreversible evolution to the Rayleigh-Jeans equilibrium distribution realizing the maximum of entropy $n^{eq}_p=T/(\beta_p - \mu)$.
Proceeding as in Refs.\cite{PRL05,PD09,nazarenko11,PR14}, the system exhibits a phase transition to condensation: For $E \le E_{\rm crit} \simeq N \beta_0 \sqrt{N_*/2}$, $\mu \to \beta_0$ and the condensate amplitude increases as the energy decreases, $n_0/N \simeq 1-(E-N \beta_0)/(E_{\rm crit}-N \beta_0)$. The fundamental mode then gets macroscopically populated $n_0 \gg n_p$, while the higher-order modes exhibit energy equipartition ${\cal E}_p^{eq} = (\beta_p-\beta_0) n_p^{eq} = T$. Note that $E_{\rm crit}$ only depends on the 
geometry of the waveguide potential, whose finite number of modes $N_*$ regularizes the ultraviolet catastrophe of classical waves.
These theoretical predictions have been confirmed by the simulations of the NLS Eq.(\ref{eq:nls_Ap}) and kinetic Eq.(\ref{eq:kin_np}): 
A quantitative agreement has been obtained without adjustable parameters even well beyond the validity regime of the kinetic equation, see Fig.~\ref{fig:2} and \cite{SI} -- we have checked through a scale-by-scale analysis \cite{biven} that even the low-order modes evolve in the weakly nonlinear regime.
As a consequence of the macroscopic population of the fundamental mode $u_0(\br)$, the  intensity pattern of the random wave $|\psi|^2(\br)$ exhibits a {\it stable self-cleaned shape} (see the movie in \cite{SI}).

{\it Impact of strong disorder.-}
For relatively large propagation lengths, a strong coupling among different modes can no longer be neglected \cite{kaminow13,mecozzi12a,mecozzi12b,mumtaz13,xiao14}.
We have extended the above theory by considering a general form of the $N_* \times N_*$ matrix ${\bf W}(z)$ modelling random mode coupling. 
The theory shows that strong disorder introduces an additional term in the kinetic Eq.(\ref{eq:kin_np}):
\begin{eqnarray}
\partial_z n_p = \Delta \bbeta_{sd} \sum_{q}  \Gamma_{pq}  {\hat {\cal R}}\big((\beta_p-\beta_q) l_{\beta}\big) \big( n_q(z) - n_p(z)\big),
\label{eq:eq_kin_strong}
\end{eqnarray}
where the matrix $\Gamma_{pq}$ describes mode coupling 
and $\Delta \bbeta_{sd}$ the corresponding amount of strong disorder \cite{SI}. 
Note that Eq.(\ref{eq:eq_kin_strong}) has a form similar to power coupling models \cite{kaminow13}.
In principle, the extra term (\ref{eq:eq_kin_strong}) breaks the conservation of  energy $E$, so that a $H-$theorem of the complete kinetic Eq.(\ref{eq:kin_np}) and (\ref{eq:eq_kin_strong}) would describe an irreversible thermalization toward an equilibrium state of power equipartition {\it among all the modes}.
Strong disorder would then deteriorate the condensation process.
However, the key observation is that mode coupling among non-degenerate modes is quenched by the Fourier transform of the correlation function of the fluctuations ${\hat {\cal R}}\big(\beta_0 l_{\beta}\big) \simeq 0$ \cite{kaminow13}, because $\beta_0 l_\beta \gg 1$ in the experiments of beam self-cleaning.
Mode coupling is then restricted to degenerate modes (${\hat {\cal R}}(0)=1$), 
and Eq.(\ref{eq:eq_kin_strong}) describes an exponential relaxation to an {\it equipartition of power within groups of degenerate modes}.
Interestingly, this is a property of the Rayleigh-Jeans distribution ($n_p^{eq}$ only depends on $\beta_p$).
This reveals that strong disorder does not deteriorate condensation, but instead {\it enforces the thermalization} to the Rayleigh-Jeans equilibrium distribution.
However, given the short fiber lengths considered in beam cleaning experiments ($\sim$10-20m) \cite{krupa17}, such an acceleration of thermalization is negligible with respect to the dramatic acceleration due to weak disorder, see Eq.(\ref{eq:ord_dis}).



\begin{figure}[]
\begin{center}
\includegraphics[width=1\columnwidth]{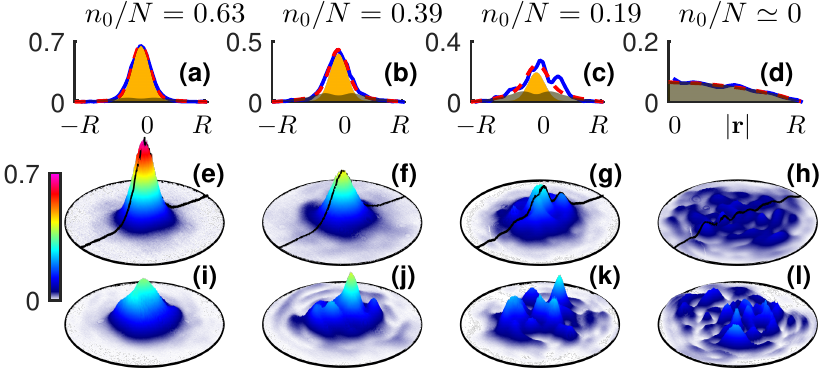}
\caption{\textbf{Experiments}.
Intensity patterns recorded at $Z_0 = 20$cm (bottom line), and at the output of the 
whole fiber length $L=13$m (middle line): By reducing the number of modes initially excited (i.e., by decreasing $E$), a transition occurs from $n_0 = 0$ to a condensate fraction of $n_0/N \simeq 0.63$.
(a-d) The fundamental Gaussian mode $u_0(\br)$ of the MMF gets macroscopically populated as evidenced by the intensity line-outs $I(x,y=0)$ recorded experimentally at $z=L$ (blue line), and the corresponding fits of the condensate contribution $I_{cond}(\br)$ (orange filled region), and of the incoherent contribution $I_{inc}(\br)$ (grey filled region). The red-dashed line denotes $I_{cond}(\br)+I_{inc}(\br)$.
Above the transition to condensation $n_0=0$ (d), we computed the angle averaged distance $|\br|$ of the intensity recorded experimentally (blue), which is in agreement with the Rayleigh-Jeans distribution $I_{inc}(\br)$ (red-dashed).
$R=26\mu$m is the fiber core radius (corresponding to the circles (e-l)), $N_*=120$ modes, $N=19$kW.
}
\label{fig:exp}
\end{center}
\end{figure}

{\it Experiments.-}
We performed experiments in a MMF to evidence the transition to light condensation by varying the coherence of the input beam. The energy $E$ provides a measure of the `coherence' of the beam in the sense that $E$ increases as the beam populates higher order modes: By increasing the coherence, 
$E$ decreases and this leads to an increase of the condensate amplitude $n_0$ after nonlinear propagation in the MMF, as described by the condensation curve in Fig.~\ref{fig:2}(c).
The sub-nanosecond pulses delivered by a Nd:YAG laser ($\lambda=1.06 \mu$m) are passed  through a diffuser to generate a beam with different properties of coherence.
The beam is subsequently injected into a graded-index MMF, and the near field intensity is recorded at the fiber output with a (CMOS) camera.
The specific fiber launch conditions are known to significantly alter the number of modes 
excited at the fiber input.
Hence, it is important to compare with the same launch conditions and the same power, the intensity pattern after a small propagation length ($Z_0$) representing the `input' (initial) field, with the corresponding intensity pattern after propagation through the whole fiber length $L$.
To this aim, the field intensity has been recorded at the output of the fiber length $L=13$m, which has been subsequently cut to $Z_0 = 20$cm to record the corresponding `input' field.

The input and output intensity patterns are reported in Fig.~\ref{fig:exp} for a power fixed to $N=19$kW (at the fiber output).
The incoherent beam evolves in the weakly nonlinear regime, $L_{lin} (\simeq 0.2$mm) $\ll L_{nl}(\simeq 1$m).
When a large number of modes are excited by the `initial' beam (i.e., $E > E_{\rm crit}$), the intensity distribution tends to relax to the equilibrium Rayleigh-Jeans distribution $I_{inc}(\br)=\sum_{p} n_p^{eq} |u_p(\br)|^2$, see Fig.~\ref{fig:exp}(d). 
By reducing the excitation of modes (i.e., $E<E_{\rm crit}$), the power gradually condenses into the fundamental Gaussian mode $u_0(\br)$ of the parabolic potential $V(\br)$, see the orange-filled region in Fig.~\ref{fig:exp}(a-c). 
This allowed us to compute with accuracy the condensate contribution in the fundamental mode $I_{cond}(\br)=n_0 u_0^2(\br)$ (orange region), and the incoherent contribution from all other modes $I_{inc}(\br)=\sum_{p\neq 0} n_p^{eq} |u_p(\br)|^2$ (grey region).
We observe a transition from a vanishing $n_0$ to a  condensate fraction of up to $\simeq 60\%$ as the coherence of the `input' beam is increased -- these measurements being only weakly affected by the Raman effect \cite{SI}.

{\it Conclusion.-}
We have shown that the previously unrecognized process of disorder-induced acceleration of condensation can explain the phenomenon of optical beam self-cleaning. 
The discrete kinetic Eq.(\ref{eq:kin_np}) also explains why beam self-cleaning has not been observed in step-index optical fibers (i.e., homogeneous potential $V(r)$) because of the absence of exact resonances.
The theory and the experiment can be extended to study turbulence cascades \cite{nazarenko11,Newell_Rumpf}, or spatio-temporal effects \cite{wright15,liu16,krupa16,wright16,laegsgaard18}.
Even more importantly, the theoretical approach developed to tackle the impact of a structural disorder is general and can be applied to different types of {\it disordered nonlinear systems} (Bose-Einstein condensates, hydrodynamics, condensed matter...).
At variance with recent observations of superfluid light flows \cite{vocke16,michel18}, our experiment of light condensation could demonstrate a key manifestation of superfluidity, namely the nucleation of vortices induced by a rotating confining potential (along the `time' $z-$variable) in manufactured fibers, in complete analogy with rotating trapped Bose-Einstein condensates \cite{stringari}.

{\it Acknowledgements.}   
The authors are grateful to A. Tonello, C. Michel, P. B\'ejot, S. Gu\'erin, V. Couderc, A. Barth\'elemy and S. Rica for fruitful discussions. We  acknowledge  financial  support  from: iXcore  research  foundation; EIPHI Graduate School (contract "ANR-17-EURE-0002"); French   program ``Investissement d'Avenir",  project  ISITE-BFC-299 (ANR-15 IDEX-0003); H2020 Marie Sklodowska-Curie Actions (MSCA-COFUND) (MULTIPLY project, no. 713694). Calculations were performed using HPC resources from DNUM CCUB (Centre de Calcul, Universit\'e de Bourgogne).



\begin{thebibliography}{99}

\bibitem{stringari}
S. Pitaevskii, L. Stringari, 
{\it Bose-Einstein Condensation} (Oxford Science Publications, 2003).



\bibitem{carusotto13}
I. Carusotto, C. Ciuti, 
Quantum fluids of light, 
Rev. Modern Phys. {\bf 85} 299 (2013).


\bibitem{weitz} 
J. Klaers, J. Schmitt, F. Vewinger, M. Weitz, 
Bose-Einstein condensation of photons in an optical microcavity,
Nature {\bf 468}, 545 (2010).


\bibitem{Newell01} 
A.C. Newell, S. Nazarenko, L. Biven, 
Wave turbulence and intermittency, 
Physica D  {\bf 152}, 520 (2001).


\bibitem{nazarenko11} 
S. Nazarenko, {\it Wave Turbulence} (Springer, Lectures Notes in Physics, 2011).

\bibitem{Newell_Rumpf} 
A.C. Newell, B. Rumpf, 
Wave Turbulence,
Annu. Rev. Fluid Mech. {\bf 43}, 59 (2011).


\bibitem{PRL05} 
C. Connaughton, C. Josserand, A. Picozzi, Y. Pomeau, S. Rica, 
Condensation of classical nonlinear waves,
Phys. Rev. Lett. {\bf 95}, 263901 (2005).



\bibitem{berloff07}
N.G. Berloff, A.J. Youd,
Dissipative dynamics of superfluid vortices at nonzero temperatures,
Phys. Rev. Lett. {\bf 99}, 145301 (2007).

\bibitem{PD09} 
G. D\"uring, A. Picozzi, S. Rica,
Breakdown of weak-turbulence and nonlinear wave condensation,
Physica D {\bf 238}, 1524 (2009).


\bibitem{Fleischer} C. Sun, S. Jia, C. Barsi, S. Rica, A. Picozzi, J. Fleischer, 
Observation of the kinetic condensation of classical waves,
Nature Phys. {\bf 8}, 471 (2012).

\bibitem{suret} 
P. Suret and S. Randoux, Far field measurement in the focal
plane of a lens: a cautionary note, arXiv:1307.5034.


\bibitem{magnons15}
A. R\"uckriegel, P. Kopietz,
Rayleigh-Jeans condensation of pumped magnons in thin-film ferromagnets,
Phys. Rev. Lett. {\bf 115}, 157203 (2015).

\bibitem{nazarenko16}
S. Nazarenko, M. Onorato, D. Proment, 
Bose-Einstein condensation and Berezinskii-Kosterlitz-Thouless transition in the two-dimensional nonlinear Schr\"odinger model,
Phys. Rev. A {\bf 90}, 013624 (2014).

\bibitem{PRL18}
N. Santic, A. Fusaro, S. Salem, J. Garnier, A. Picozzi, R. Kaiser,
Nonequilibrium precondensation of classical waves in two dimensions propagating through atomic vapors,
Phys. Rev. Lett. {\bf 120}, 055301 (2018).

\bibitem{zakharov92}
V.E. Zakharov, V.S. L'vov, G. Falkovich, 
{\it Kolmogorov Spectra of Turbulence I} (Springer, Berlin, 1992).

\bibitem{shrira_nazarenko13}
{\it Advances in Wave Turbulence,} World Scientific Series on
Nonlinear Science Series A, Vol. 83, edited by V.I. Shrira
(World Scientific, Singapore, 2013).




\bibitem{chiocchetta16} 
A. Chiocchetta, P.E. Larr\'e, I. Carusotto, 
Thermalization and Bose-Einstein condensation of quantum light in bulk nonlinear media,
Europhys. Lett. {\bf 115}, 24002 (2016).

\bibitem{PR14}
A. Picozzi, J. Garnier, T. Hansson, P. Suret, S. Randoux, G. Millot,  D.N. Christodoulides, 
Optical wave turbulence: Toward a unified nonequilibrium thermodynamic formulation of statistical nonlinear optics,
Physics Reports {\bf 542}, 1-132 (2014).

\bibitem{laurie12}
J. Laurie, U. Bortolozzo, S. Nazarenko, S. Residori, 
One-dimensional optical wave turbulence: experiment and theory, 
Physics Reports {\bf 514}, 121-175 (2012).


\bibitem{vocke16}
D. Vocke, K. Wilson, F. Marino, I. Carusotto, E.M. Wright, T. Roger, B.P. Anderson, P. \"Ohberg, D. Faccio,
Role of geometry in the superfluid flow of nonlocal photon fluids,
Phys. Rev. A {\bf 94}, 013849 (2016).

\bibitem{michel18}
C. Michel, O. Boughdad, M. Albert, P.-E. Larr\'e, M. Bellec,
Superfluid motion and drag-force cancellation in a fluid of light,
Nature Comm. {\bf 9}, 2108 (2018).



\bibitem{conti08}
C. Conti, M. Leonetti, A. Fratalocchi, L. Angelani, G. Ruocco, 
Condensation in disordered lasers: Theory, 3d+1 simulations, and experiments, 
Phys. Rev. Lett. {\bf 101},  143901 (2008).


\bibitem{fischer14}
G. Oren, A. Bekker, B. Fischer, 
Classical condensation of light pulses in a loss trap in a laser cavity,
Optica {\bf 1}, 145 (2014).



\bibitem{turitsyn13}
E. Turitsyna, S. Smirnov, S. Sugavanam, N. Tarasov, X. Shu, S. Babin, E. Podivilov, D. Churkin, G. Falkovich, S. Turitsyn, 
The laminar-turbulent transition in a fibre laser, 
Nature Photon. {\bf 7}, 783 (2013).




\bibitem{krupa16}
K. Krupa, A. Tonello, A. Barth\'el\'emy, V. Couderc, B.M. Shalaby, A. Bendahmane, G. Millot, S. Wabnitz, 
Observation of geometric parametric instability induced by the periodic spatial self-imaging of multimode waves,
Phys. Rev. Lett. {\bf 116}, 183901 (2016).

\bibitem{wright16}
L.G. Wright, Z. Liu, D.A. Nolan, M.-J. Li, D.N. Christodoulides, F.W. Wise,
Self-organized instability in graded-index multimode fibres,
Nature Photon. {\bf 10}, 771 (2016).

\bibitem{krupa17}
K. Krupa, A. Tonello, B.M. Shalaby, M. Fabert, A. Barth\'el\'emy, G. Millot, S. Wabnitz, V. Couderc,
Spatial beam self-cleaning in multimode fibres,
Nature Photon. {\bf 11}, 237 (2017).


\bibitem{kaminow13} 
I.P. Kaminow, T. Li, A.F. Willner, 
{\it Optical Fiber Telecommunications, Systems and Networks} (Sixth Ed., Elsevier, 2013).

\bibitem{psaltis16}
D. Psaltis, C. Moser,
Imaging with multimode fibers,
Opt. and Photon. News {\bf 27}, 24 (2016).

\bibitem{mecozzi12a}
A. Mecozzi, C. Antonelli, M. Shtaif, 
Nonlinear propagation in multimode fibers in the strong coupling regime,
Opt. Exp. {\bf 20}, 11673 (2012).

\bibitem{mecozzi12b}
A. Mecozzi, C. Antonelli, M. Shtaif, 
Coupled Manakov equations in multimode fibers with strongly coupled groups of modes, 
Opt. Exp. {\bf 20}, 23436 (2012).

\bibitem{mumtaz13}
S. Mumtaz, R.J. Essiambre, G.P. Agrawal, 
Nonlinear propagation in multimode and multicore fibers: Generalization of the Manakov
equations,
J. Lightw. Technol. {\bf 31}, 398 (2013).

\bibitem{xiao14}
Y. Xiao, R.-J. Essiambre, M. Desgroseilliers, A.M. Tulino, R. Ryf, S. Mumtaz, G.P. Agrawal,
Theory of intermodal four-wave mixing with random linear mode coupling in few-mode fibers,
Opt. Exp. {\bf 22}, 32039 (2014).




\bibitem{Lvov10}
V.S. L'vov, S.V. Nazarenko,
Discrete and mesoscopic regimes of finite-size wave turbulence,
Phys. Rev. E {\bf 82}, 056322 (2010).

\bibitem{Harris13}
J. Harris, C. Connaughton, M.D. Bustamante,
Percolation transition in the kinematics of nonlinear resonance broadening in Charney-Hasegawa-Mima model of Rossby wave turbulence,
New Journal of Physics {\bf 15}, 083011 (2013).

\bibitem{Harper13}
K.L. Harper, M.D. Bustamante, S.V. Nazarenko,
Quadratic invariants for discrete clusters of weakly interacting waves,
J. Phys. A: Math. Theor. {\bf 46}, 245501 (2013).


\bibitem{Bustamante14}
M. Bustamante, B. Quinn, D. Lucas, 
Robust Energy Transfer Mechanism via Precession Resonance in Nonlinear Turbulent Wave Systems,
Phys. Rev. Lett. {\bf 113}, 084502 (2014).













\bibitem{Ruffo2009} 
A. Campa, T. Dauxois, S. Ruffo, 
Statistical mechanics and dynamics of solvable models with long-range interactions,
Physics Reports {\bf 480}, 57 (2009).

\bibitem{Schmiedmayer16} 
T. Langen, T. Gasenzer, J. Schmiedmayer,
Prethermalization and universal dynamics in near-integrable quantum systems,
J. Stat. Mech. {\bf 6}, 064009 (2016).





\bibitem{leuzzi15}
F. Antenucci, M. Ibanez Berganza, L. Leuzzi,
Statistical physics of nonlinear wave interaction, 
Phys. Rev. B {\bf 92}, 014204 (2015).




\bibitem{delre17}
D. Pierangeli, A. Tavani, F. Di Mei, A.J. Agranat, C. Conti, E. DelRe,
Observation of replica symmetry breaking in disordered nonlinear wave propagation,
Nature Commun. {\bf 8}, 1501 (2017).


\bibitem{churkin15}
D. Churkin, I. Kolokolov, E. Podivilov, I. Vatnik, S. Vergeles, I. Terekhov, 
V. Lebedev, G. Falkovich, M. Nikulin, S. Babin, S. Turitsyn, 
Wave kinetics of a random fibre laser, 
Nature Commun. {\bf 2}, 6214 (2015).


\bibitem{silberberg17}
H. Frostig, E. Small, A. Daniel, P. Oulevey, S. Derevyanko, Y. Silberberg,
Focusing light by wavefront shaping through disorder and nonlinearity,
Optica {\bf 4}, 1073 (2017).

\bibitem{segev}
M. Segev, Y. Silberberg, D.N. Christodoulides, 
Anderson localization of light,
Nature Photonics {\bf 7}, 197 (2013).  

\bibitem{cherroret15}
N. Cherroret, T. Karpiuk, B. Gr\'emaud, C. Miniatura,
Thermalization of matter waves in speckle potentials,
Phys. Rev. A {\bf 92}, 063614 (2015).

\bibitem{schirmacher18}
W. Schirmacher, B. Abaie, A. Mafi, G. Ruocco, M. Leonetti,
What is the right theory for Anderson localization of light? An experimental test,
Phys. Rev. Lett. {\bf 120}, 067401 (2018).


\bibitem{ermann15}
L. Ermann, E. Vergini, D.L. Shepelyansky,
Dynamical thermalization of Bose-Einstein condensate in Bunimovich stadium,
Europhys. Lett. {\bf 111}, 50009 (2015). 





\bibitem{wright17}
L.G. Wright, D.N. Christodoulides, F.W. Wise,
Spatiotemporal mode-locking in multimode fiber lasers,
Science {\bf 358}, 94 (2017).



\bibitem{horak12}
P. Horak, F. Poletti, 
Multimode nonlinear fibre optics: Theory and applications, 
in {\it Recent Progress in Optical Fiber Research}, 
M. Yasin, S. W. Harun, and H. Arof, eds. (InTech, 2012), 3-25.


\bibitem{SI}
See Supplemental Material [url] for a movie, for the derivation of the kinetic Eq.(3-5), and the analytical expression of the condensate fraction in Fig.~2(c), for a detailed comparison of the simulations with the kinetic theory (scaling of acceleration of thermalization in Eq.(4)), and for a complementary description of the experimental methods (e.g. measurements of the condensate fraction in Fig.~3), which includes Refs.[51$-$56].











\bibitem{menyuk}
P. Wai, C.R. Menyuk, 
Polarization mode dispersion, decorrelation, and diffusion in optical fibers with randomly varying birefringence, 
J. Lightw. Technol. {\bf 14}, 148 (1996).






\bibitem{mecozzi12c}
C. Antonelli, A. Mecozzi, M. Shtaif, P.J. Winzer, 
Stokes-space analysis of modal dispersion in fibers with multiple mode transmission,
Opt. Exp. {\bf 20}, 11718 (2012).




\bibitem{PRA11b}
P. Aschieri, J. Garnier, C. Michel, V. Doya, A. Picozzi,
Condensation and thermalization of classsical optical waves in a waveguide,
Phys. Rev. A {\bf 83}, 033838 (2011).


\bibitem{kohler}
{W. Kohler and G. Papanicolaou, Wave Propagation in Randomly Inhomogeneous Ocean, J.B. Keller and J.S. Papadakis, eds., 
{\it Wave Propagation and Underwater Acoustics} (Springer-Verlag, Berlin, LNP 70, 1977).}

\bibitem{laegsgaard17}
J. Laegsgaard,
Efficient simulation of multimodal nonlinear propagation in step-index fibers,
J. Opt. Soc. Am. B {\bf 34}, 2266 (2017).


\bibitem{zhan09}
Y. Zhan, Q. Yang, Hua Wu, J. Lei, P. Liang,
Degradation of beam quality and depolarization of the laser beam in a step-index multimode optical fiber,
Optik {\bf 120}, 585 (2009). 




















\bibitem{biven}
L. Biven, S.V. Nazarenko, A.C. Newell, 
Breakdown of wave turbulence and the onset of intermittency, 
Phys. Lett. A {\bf 280}, 28 (2001).


\bibitem{wright15}
L.G. Wright, S. Wabnitz, D.N. Christodoulides, F.W. Wise,
Ultrabroadband dispersive radiation by spatiotemporal oscillation of multimode waves,
Phys. Rev. Lett. {\bf 115}, 223902 (2015).







\bibitem{laegsgaard18}
J. Laegsgaard,
Spatial beam cleanup by pure Kerr processes in multimode fibers,
Opt. Lett. {\bf 43}, 2700 (2018).

\bibitem{liu16}
Z. Liu, L.G. Wright, D.N. Christodoulides, F.W. Wise, 
Kerr self-cleaning of femtosecond-pulsed beams in graded-index multimode fiber,
Optics Letters {\bf 41} 3675  (2016).






















































\end{thebibliography}

\end{document}